\documentstyle[12pt]{article}
\newcommand{\f}{\begin{equation}}
\newcommand{\ff}{\end{equation}}
\newcommand{\blankmm}{\vskip .2cm}
\newcommand{\blankline}{\vskip .3cm}
\newcommand{\blankblock}{\vskip 3cm}

\newcommand{\half}{{1\over2}}

\newcommand{\pl}{Phys. Lett.\ }

\newcommand{\np}{Nucl. Phys.\ }

\newcommand{\ep}{\epsilon}

\def\beq{\begin{equation}}
\def\beqa{\begin{eqnarray}}
\def\eeq{\end{equation}}
\def\eeqa{\end{eqnarray}}
\def\A{{\cal A}}
\def\G{{\cal G}}
\def\Emi{E_{-\alpha_i}}
\def\Epl{E_{\alpha_i}}
\def\a{\alpha}
\def\H{{\cal H}}

\def\lra{\big\vert\lambda_i\big>}
\def\lrab{\big\vert\lambda_{\bar i}\big>}
\def\lla{\big<\lambda_i\big\vert}
\def\llab{\big<\lambda_{\bar i}\big\vert}

\def\gte{\geq}
\def\vu{\nu}

\def\aE{\a E}

\begin{document}
\begin{titlepage}

\begin{flushright}
UM-P-95/31 \\
QMW-PH-95/7 \\
hep-th/9505026 \\
March 1995 \\
\end{flushright}

\blankblock
\centerline{\LARGE Hamiltonian Reduction and Supersymmetric }
\blankmm
\centerline{\LARGE Toda Models}
\blankline
\blankline
\blankline
\centerline{G. Au\footnote{gau@tauon.ph.unimelb.edu.au}
     and B. Spence\footnote{B.Spence@qmw.ac.uk. Current address:
   Dept. Physics, Queen Mary \& Westfield College, London E1 4NS.}}
\centerline{School of Physics}
\centerline{University of Melbourne}
\centerline{Parkville 3052}
\centerline{Australia}
\blankline

\begin{abstract}
New formulations of the solutions of N=1 and N=2 super Toda field theory
are introduced, using Hamiltonian reduction of the N=1 and N=2 super WZNW
Models to the super Toda Models.
These parameterisations are then used to present the Hamiltonian formulations
 of the super Toda theories on the spaces of solutions.
\end{abstract}
\vfill
\end{titlepage}

\newpage
\section{Introduction}
Super Toda theories \cite{lss,nm,eh,ms,grisaru}
are two dimensional conformal field theories and are an arena
for the investigation of general features of two dimensional
 integrable systems such as
quantum supergroups, super W-algebras, and supersymmetric
integrable hierachies.
The study of many of these properties requires
Hamiltonian formulations of the models. The Toda theories can be understood as
the
Hamiltonian reductions of Wess-Zumino-Novikov-Witten (WZNW)
models (see ref. \cite{bfof}).
As
each WZNW model is integrable, one
can explicitly write a Hamiltonian formulation in terms
of the coordinates of the space of solutions of the model. In order to preserve
full equivalence with the Hamiltonian formulation using the
space of initial data, one needs to formulate the
WZNW solutions as discussed in ref. \cite{pswzw} (see also ref.
\cite{mansfield}).
Using this formulation, one may derive,
by Hamiltonian reduction, a Hamiltonian formulation of
Toda theory \cite{pstoda},
utilising suitable coordinates on the space of solutions.
It is natural to seek to formulate the supersymmetric analogues of these
results.

This would be most easily done in superspace. However, full superspace
realisations
of generic supersymmetric WZNW models are only
available for the $(1,0)$ and $(1,1)$ models,
and certain $N\geq2$ models (for
$(2,0)$ and $(2,1)$ see ref. \cite{hs},
whilst for $(2,2)$ and $N>2$ see ref. \cite{kr}
for a recent discussion and references).
In those cases where a superspace formulation is not available, one can choose
to
work in some suitable component formulation. By \lq$N=k$'
we will mean $(k,k)$ supersymmetry ($k=1$ or $2$) in the following.

The Hamiltonian formulation of the WZNW models of ref. \cite{pswzw}
was recently generalised to the
$N=1$ and some $N=2$ super WZNW models in ref. \cite{as}.
In this paper, we will show how
one may apply a Hamiltonian reduction procedure to these super WZNW models to
obtain new Hamiltonian formulations of the super Toda theories.

In section 2 we will review the formulation of Hamiltonian
reduction as applied to $N=1$ super WZNW models, and will show how this
may be done for the $N=2$ models (for the $N=2$ Liouville model, this was
done in ref. \cite{soh}).
Then, in section 3, we will show how this procedure applied to the super WZNW
solutions presented in ref.
\cite{as} yields new formulations
of the solutions of super
Toda theories. In section 4, we will use these new phase
space coordinate systems to write
Hamiltonian formulations of the super Toda
theories. We finish with some concluding remarks in section 5.

\section {Super Toda Field Theory and Hamiltonian Reduction}

\subsection {$N=1$ Hamiltonian Reduction}

Here we will follow the Lie super-algebra conventions of ref. \cite{drs}.
The superalgebra $\A$ is one of
the set $sl(n+1|n), Osp(2n\pm 1|2n), Osp(2n|2n), Osp(2n+2|2n), n\gte 1$, or
$D(2,1;\a), \a\not=0,-1$. In the Chevalley basis we have Cartan
generators $H_i, i=1,...,r={\rm rank}(\A)$, and
fermionic generators $E_{\pm\alpha_i}$
associated to the simple root system $(\alpha_1,...,\alpha_r)$.
$K_{ij}$ is the Cartan matrix. $\A_+ (\A_-)$ denotes the superalgebra spanned
by the
generators of $\A$ corresponding to the positive (negative) roots respectively,
with
respect to the Cartan subalgebra $\H$. $\G_+, \G_-, \G$ are
the corresponding groups associated to the algebras $\A_+, \A_-, \A$.

The $N=1$ super WZNW model has the equations of motion
\begin{equation}\label{toda1eqns}
D_-(D_+G G^{-1}) = 0
\end{equation}
for $G\in\G$. Our superspace has coordinates $(x^{\pm\pm}, \theta^\pm)$ and we
use
the supercovariant derivatives $D_\pm = \partial/\partial\theta^\pm +
\theta^\pm
\partial_{\pm\pm}$. We will work in Minkowski space-time, with coordinates
$(t,x)$ and signature $(-1,1)$, and take
$x^{\pm\pm} = t\pm x, \partial_{\pm\pm}=(1/2)(\partial_t\pm\partial_x).$
We now impose the following constraints
\beqa\label{wzwcurrents}
(D_+GG^{-1})|_- &= \kappa \mu_i\Emi,\nonumber\\
(G^{-1}D_-G)|_+ &= \kappa \nu_i\Epl,
\eeqa
for $\alpha_i$ simple, where $|_\pm$ means the components along the
positive (negative) root systems, $\mu_i$ and $\nu_i$ are
constants and $\kappa$ is a coupling constant.
Then, as discussed in ref. \cite{drs},
under the constraints (\ref{wzwcurrents})
the $N=1$ super WZNW field equations (\ref{toda1eqns})
reduce to the $N=1$ super Toda equations of motion
\begin{equation}\label{toda}
D_+D_-\Phi^i  -\kappa^2\mu_i\vu_ie^{K_{ij}\Phi^j} = 0, \quad (i=1,...,r),
\end{equation}
(the repeated index $i$ is not summed over in this or later equations of
motion).
As we will later
study the $N=2$ models in
components, it is
instructive to see how this works for the $N=1$ case. The basic fields
are a bosonic $g\in\G$ and fermions
$\psi_\pm\in{\rm Lie}\G$, the Lie algebra of $\G$. These
are given by $g=G|_0, \psi_+=(D_+G\,G^{-1})|_0$ and
$\psi_-=(G^{-1}D_-G)|_0$, where $|_0$ means to evaluate at $\theta^\pm=0$.
The equations of
motion of the $N=1$ super WZNW model are $\partial_{--}(\partial_{++}g\,
g^{-1})=0,
\partial_{--}\psi_+=\partial_{++}\psi_-=0$.
The (on-shell) $N=1$ supersymmetry transformations are
\begin{eqnarray}\label{neqone}
\delta^+g &=&\ep^+\psi_+g, \quad\quad \delta^-g = \ep^-g\psi_- \nonumber\\
\delta^+\psi_+ &=& \ep^+(\partial_{++}g g^{-1}+ \psi^2_+),
\quad \delta^-\psi_-
= \ep^-(g^{-1}\partial_{--}g  - \psi^2_-)\nonumber\\
\delta^+\psi_- &=& 0, \quad\quad\quad\quad \delta^-\psi_+ = 0.
\end{eqnarray}
The Hamiltonian reduction constraints are the fermionic constraints
\beqa\label{oneoneconstraints}
\psi_+|_- &= \kappa \mu_i\Emi,\nonumber\\
\psi_-|_+ &= \kappa \nu_i\Epl,
\eeqa
and their supersymmetric partners
\beqa\label{moreoneone}
(\partial_{++}g\, g^{-1} + \psi_+^2)|_- &= 0,\nonumber \\
(g^{-1}\partial_{--}g - \psi_-^2)|_+ &= 0.
\eeqa
Under these constraints, the super WZNW equations of motion reduce to the $N=1$
super
Toda equations of motion
\beqa\label{oneonetoda}
\partial_{++}\partial_{--} \phi^i &=&
 - M_i(K_{ij}\chi_+^j)(K_{ik}\chi_-^k)\exp(K_{il}\phi^l)\nonumber\\
  &&  \quad   -M_iM_jK_{ij}\exp(K_{jk}\phi^k)\exp(K_{il}\phi^l)  \nonumber\\
\partial_{++}\chi_-^i &=& -M_iK_{ik}\chi_+^k\exp(K_{ij}\phi^j) \nonumber\\
\partial_{--}\chi_+^i &=& M_iK_{ik}\chi_-^k\exp(K_{ij}\phi^j),
\eeqa
where $M_i = \kappa^2\mu_i\nu_i$ and
$\chi_\pm^i$ are the supersymmetric partners of $\phi^i$.
In terms of the superfield $\Phi^i$ of equation (\ref{toda}),
these fields are given by $\phi^i = \Phi^i|_0$, $\chi_\pm^i=D_\pm\Phi^i|_0$.
The auxiliary field $F^i=D_+D_-\Phi^i|_0$ is eliminated using its
(non-propagating) equation of motion in the usual way.


\subsection {$N=2$ Hamiltonian Reduction}

There is currently no off-shell
superfield formulation of generic $N=2$ supersymmetric WZNW models, and we
will
thus work in components. The $N=2$ supersymmetry
requires the supergroup to be
$sl(n+1|n)$, for some positive integer $n$. It proves convenient to use the
left and right invariant one forms on the group manifold, $L=g^{-1} dg,
R= dg\, g^{-1}$,
satisfying $dR = R^2, dL = -L^2$, or in components
\beq\label{blah}
R^A_{[I,J]} = -\half {f^A}_{BC} R^B_I R^C_J,\qquad
L^A_{[I,J]} = \half {f^A}_{BC}L^B_I L^C_J,
\eeq
where all the indices run from $1$ to ${\rm dim}\G$, with $A,B,C,...,$ Lie
algebra
indices and $I,J,K,...,$ coordinate indices. The vector fields
$R_A = R_A^I\partial_I, L_A = L_A^I\partial_I$ ($\partial_I=\partial/\partial
X^I$,
where $X^I$ are the group manifold coordinates) also satisfy
\beq\label{blahh}
[R^A,R^B]=-{f^{AB}}_C R^C,\quad [L^A,L^B] = {f^{AB}}_C L^C,\quad [R^A,L^B] = 0.
\eeq
Finally we have $R^A_IR^I_B = \delta^A_B, R^A_IR^J_A=\delta^J_I$, with
similar equations for $L$.

Under the left and right complex structures on the group manifold, the
coordinates naturally split
into sets with barred or unbarred indices
$I\rightarrow(i,\bar i), A\rightarrow(a,\bar a)$, etc. Furthermore we have
the properties $f_{abc}=0=K_{ab}$ and their conjugates.

The $N=2$ super WZNW model has the bosonic field
$g\in\G$, and fermions $\psi^a_{\pm},
\bar\psi_\pm^{\bar a} \in {\rm Lie}\G$, with equations of motion
\beq\label{twotwoeqm}
\partial_{--}(\partial_{++}g g^{-1}) = 0,\qquad
\partial_{\pm\pm}\psi_\mp^a = 0 = \partial_{\pm\pm}\bar\psi^{\bar a}_\mp.
\eeq
The right supersymmetry transformations are
\beqa\label{twotwosusy}
\delta X^I&=& \ep^+R^I_a\psi_+^a + \bar\ep^+R^I_{\bar a}\bar\psi_+^{\bar a},
\nonumber\\
\delta \psi_+^a &=& \half\ep^+ {f^a}_{bc}\psi^b_+\psi_+^c + \bar\ep^+
     {f^a}_{b\bar c}\psi^b_+\bar\psi^{\bar c}_+ +
\bar\ep^+R^a_I\partial_{++}X^I,
\nonumber\\
\delta\bar\psi^{\bar a}_+ &=& \half\bar\ep^+{f^{\bar a}}_{\bar b\bar
c}\bar\psi_+
^{\bar b}\bar\psi_+^{\bar c} + \ep^+{f^{\bar a}}_{b\bar c}\psi^b_+
\bar\psi_+^{\bar c}
+\ep^+R^{\bar a}_I\partial_{++}X^I, \nonumber\\
\delta\psi_-^a &=& \delta\bar\psi_-^{\bar a} =0,
\eeqa
with the left supersymmetry transformations being given by the above formul\ae\
with the
replacements $\psi_+\rightarrow\psi_-, \partial_{++}\rightarrow\partial_{--},
f_{ABC}\rightarrow-f_{ABC}, R^A_I\rightarrow L^A_I$.

The natural fermionic
Hamiltonian reduction constraints to take are
\beqa\label{twotwofcon}
\psi_+|_{-} = \kappa \mu^a E_-^a,\qquad
\psi_-|_{+} = \kappa \nu^a E_+^a, \nonumber \\
\bar{\psi}_+|_{-} = \kappa \mu^{\bar{a}}  E_-^{\bar{a}},\qquad
\bar{\psi}_-|_{+} = \kappa \nu^{\bar{a}}  E_+^{\bar{a}},
\eeqa
where $|_{\pm}$ means the components along the positive or negative root
systems. $E_\pm^a, E_\pm^{\bar a}$ are the fermionic step operators
corresponding to the simple roots.
We also impose the supersymmetry partners of these constraints,
which are
\beqa\label{twotwosusypart}
\psi_+^2 |_{-} = 0, \qquad \psi_-^2 |_{+} = 0, \nonumber \\
\bar{\psi}_+^2 |_{-} = 0, \qquad
\bar{\psi}_-^2 |_{+} = 0, \nonumber \\
(\partial_{++}g\,g^{-1} + \{ \psi_+, \bar{\psi}_+\}) | _{-} = 0,
\nonumber \\
(g^{-1}\partial_{--}g - \{ \psi_-, \bar{\psi}_-\}) | _{+} = 0.
\eeqa
Under the constraints (\ref{twotwofcon}) and (\ref{twotwosusypart}),
the WZNW system (\ref{twotwoeqm}) reduces to the $N=2$ super
Toda theory whose equations of motion will be given below in Section 3.2.

We have not discussed possible Hamiltonian reductions of \lq chiral'
WZNW models, with $(p,q)$, $p\not= q$, supersymmetry. Lagrangian
formulations of these models exist, as noted in the introduction, and
 one may impose constraints to reduce these theories to systems with
Toda-type
equations of motion.
However, to our knowledge there are no Lagrangian formulations of $(p,q)$,
$p\not=q$ supersymmetric Toda theories and for
this reason we have not further investigated these chiral reductions.


\section{The Super Toda Solutions}

In the previous section it was shown how to impose constraints in order
to reduce
the $N=1$ and $2$
super WZNW models to the super Toda theories.
One may use this to reduce solutions of
super WZNW models to solutions of super Toda theories.
The former have been presented in ref. \cite{as}.
In this section,
we will show how to obtain new formulations of the
solutions of $N=1,2$ super Toda
theories by Hamiltonian reduction of these WZNW solutions.

\subsection{$N=1$ Super Toda Solutions}

For the $N=1$ model we begin with the solution of ref. \cite{pswzw} to the
field
equations (\ref{toda1eqns}), {\it i.e.}
\begin{eqnarray}\label{param}
 G(t,x,\theta^+) &=& U(x^{++},\theta^+){\em
W}(A;x^{++},x^{--})V(x^{--},\theta^-),
 \nonumber \\
 W(A;x^{++},x^{--}) &=& P \, {\rm exp}\! \int_{x^{--}}^{x^{++}}A(s) ds,
\end{eqnarray}
with $A$ a $({\rm Lie} \G)^*$-valued one-form on the real line, $P$
denoting path ordering.
The parameterisation (\ref{param}) of the solutions is
invariant under the group transformations
\begin{eqnarray}\label{transf}
 &U(x,\theta^+)& \longrightarrow U(x,\theta^+)h(x), \,\,\,
   V(x,\theta^-) \longrightarrow h^{-1}(x)V(x,\theta^-),
 \nonumber \\
 &A(x)&\longrightarrow - h^{-1}(x)\partial_x h(x) + h^{-1}(x)
  A(x)h(x),
\end{eqnarray}
where $h$ is an element of the loop group of $\G$.

Rewriting the constraints (\ref{wzwcurrents}) in terms of the variables $U,V,A$
of the parameterisation (\ref{param}), they become
\begin{eqnarray}\label{cons}
   (D_+U U^{-1}+ UA(x^{++})U^{-1})|_-
	&= \kappa\mu_i\Emi, \nonumber \\
   (V^{-1}D_-V -V^{-1}A(x^{--})V)|_+
	&= \kappa\vu_i\Epl.
\end{eqnarray}
To solve these constraints, we recall
that the group $\G$ admits (locally) a Gauss decomposition, so that
elements $G\in\G$ can be decomposed as $G=DBC$, where
$D$ and $C$ lie in $\G_+$ and $\G_-$ respectively,
and $B$ lies in a maximal torus of $\G$ with Lie algebra $\H$. The
parameters $U,V$ can thus be decomposed as
\begin{equation}\label{decomp}
	U = D_LB_LC_L, \quad V = D_RB_RC_R,
\end{equation}
for some
\begin{eqnarray}\label{decomp2}
   D_L &=& \exp(\sum_{\a\in\Pi_+}\hat X_L^\aE_\a),
     \quad C_L = \exp(\sum_{\a\in\Pi_+}\hat Y_L^\aE_{-\a}), \nonumber \\
    B_L &=& \exp(\hat\Phi^i_L H_i),
\end{eqnarray}
($\Pi_+$ is the set of positive roots)
and similarly for $V$, with left subscripts
$L$ replaced by right subscripts $R$.
The constraints (\ref{cons}) can be rewritten in terms
of  $D_R$, $\hat\Phi_L$, $\hat\Phi_R$ and $C_L$  as follows:
\begin{eqnarray}\label{cons2}
(D_+C_L\,C_L^{-1} + \theta^+ C_LA(x^{++})C_L^{-1})|_- &=&
 \sum_{\a_i\in\Delta_+}\mu_i\Emi{\rm exp}(
             K_{ij}\hat\Phi_L^j),
\nonumber\\
(D_R^{-1}D_-D_R - \theta^-D_R^{-1}A(x^{--})D_R)|_+ &=&
 -\sum_{\a_i\in\Delta_+}\vu_i\Epl{\rm exp}(
             K_{ij}\hat\Phi_R^j),
\end{eqnarray}
with $\Delta_+$ the set of simple roots.

To find the independent
parameters of the solutions of the Toda equations, we have still to gauge-fix
the symmetry (\ref{transf}) and solve the constraints (\ref{cons2}). The
transformation (\ref{transf}) can be
gauge-fixed by setting the $\theta^+$ independent parts of $U$ to one.
Next, as the constraints (\ref{cons2})
do not contain $C_R$ and $D_L$, we can set $C_R=D_L=1$ (alternatively this can
be
understood as fixing the associated shift symmetry).
The gauge symmetry
\begin{eqnarray}
    & D_R(x^{--},\theta^-) \longrightarrow
h_+^{-1}(x^{--})D_R(x^{--},\theta^-),
\nonumber \\
    & A(s) \longrightarrow -h_+^{-1}(s)\partial_{s}h_+(s) + h_+^{-1}(s)
A(s)h_+(s),
\end{eqnarray}
where $h_+$ lies in the loop group of $\G_+$,
leaves $G$ and the constraints (\ref{cons2}) invariant.
We gauge fix this symmetry
by setting the $\theta^-$ independent component of $D_R$ to
$1$. We use a similar gauge symmetry for $C_L$ to set the $\theta^+$
independent component of $C_L$ to $1$.
Hence, after gauge-fixing we can write
\beqa\label{whadda}
D_L&=&1,\quad B_L= 1 + \theta^+\psi_+^i(x^{++})H_i,
\nonumber\\
C_L &=& 1 + \theta^+\sum_{\alpha\in\Pi_+}Y_L^{\alpha_i}(x^{++})\Emi, \quad
  D_R= 1 + \theta^-\sum_{\alpha\in\Pi_+}X_R^{\alpha_i}(x^{--})\Epl, \nonumber\\
B_R &=& {\rm exp}(\hat\Phi^i_R(x^{--},\theta^-)H_i),\quad C_R=1,
\eeqa
for some (super-)fields $\psi_+^i(x^{++}), Y_L^\alpha(x^{++}),
X_R^\alpha(x^{--}),
\hat\Phi^i_R(x^{--},\theta^-)$. Solving for these fields by substituting
the expressions (\ref{whadda}) into
the constraints (\ref{cons2}) then
leads straightforwardly to the final gauge fixed parameterisation
\begin{eqnarray}\label{doowhaddy}
 U &=& 1 + \theta^+({\psi^i_+}H_i + \kappa\mu_i\Emi), \nonumber \\
 V &=& [1 + \theta^-(\psi_-^iH_i + \kappa\nu_i \exp(K_{ij}\phi_R^j)\Epl)]
 \exp{(\phi_R^k H_k)},\nonumber\\
 A &=&\hat A\equiv a^iH_i - \kappa\mu_i(K_{ij}\psi_+^j)\Emi +
      \kappa\vu_i(K_{ik}\psi_-^k)
          \exp(K_{ij}\phi_R^j)\Epl   \nonumber \\
     &&\qquad-(\kappa\mu_i\Emi)^2 -
(\kappa\vu_i\exp(K_{ij}\phi_R^j)\Epl)^2,
\end{eqnarray}
where  we have expanded $\hat\Phi^i_R(x^{--},\theta^-)=\phi_R^i(x^{--}) +
\theta^-\psi^i_-(x^{--})$.

The field
\beq
G=U\exp(\int_{x^-}^{x^+}\!\! A(s)\, ds)V,
\eeq
with the expressions (\ref{doowhaddy}) inserted, is then the
solution of the system of equations (\ref{toda1eqns}) and (\ref{wzwcurrents}).
Using this we can directly obtain the $N=1$ super Toda
solution as follows.
We first perform a Gauss decomposition for the WZNW field $G$
\begin{equation}\label{finaldecomp}
     G = DBC,
\end{equation}
where
\begin{eqnarray}\label{finalparam}
    D &=& \exp{(\sum_{\a\in\Pi_+}x^\aE_\a)},
     \quad C = \exp{(\sum_{\a\in\Pi_+}y^\aE_{-\a})}, \nonumber \\
    B &=& \exp{(\phi^iH_i)},
\end{eqnarray}
for some fields $x^\alpha, y^\alpha, \phi^i$.
The Toda field is identified as the field $\phi$ that appears in the definition
of $B$ in (\ref{finalparam}). To project it from the expression for
$G$ in (\ref{finaldecomp}) we use the standard method of the $l$ normalised
lowest weight states $| \lambda_i \rangle$ of finite dimensional
representations
of $G$ (with $H_j\lra =-\delta_{ij}\lra)$. From equations
(\ref{finalparam}),
(\ref{param}), and $<\l_i|G|\l_i>={\rm exp}(- \Phi^i)$
we obtain the result
\begin{eqnarray}\label{bigres}
 &&   \exp(-\Phi^i(\theta^+,\theta^-,x,t))=
\exp(-\phi_R^i)
<\lambda_i| [1+\theta^+(\psi_+^j H_j + \kappa\mu_jE_{-\alpha_j})] \times
\nonumber\\
&&\quad\quad        {\cal W}(\hat A;x^{++},x^{--})
       \times\, [1 + \theta^-
(\kappa\nu_j\exp(K_{jk}\phi_R^k)E_{\alpha_j} + \psi_-^j H_j)]|\lambda_i>.
\nonumber\\
\end{eqnarray}
The free parameters of this formulation of the super Toda solutions are the
chiral fields $\psi_+^i(x^{++}), \psi_-^i(x^{--}),\phi_R^i(x^{--})$ and
the field $a^i(s)$ which appears in $\hat A(s)$ in $W(\hat A,x^{++},x^{--})$.

The solutions for the component fields of $\Phi$ are readily found from
equation
(\ref{bigres}) and
are given by
\begin{eqnarray}\label{toodleoo}
\phi^i \equiv \Phi^i|_0&=&  \phi_R^i - {\rm log}
\lla W(\hat A)\lra, \nonumber\\
\chi^i_+\equiv
D_+\Phi|_0 &=& { \lla (\psi_+\cdot H + \kappa\mu\cdot E_-)W(\hat A)\lra \over
\lla
W(\hat A)\lra},        \nonumber\\
\chi^i_-\equiv
D_-\Phi|_0 &=& { <\lambda_i |
W(\hat A)
(\psi_-\cdot H + \kappa\exp(K_{ij}\phi_R^j)\nu\cdot E_+)
\lra \over \lla W(\hat A)\lra}.
\end{eqnarray}
By construction, these
fields satisfy the equations of motion (\ref{oneonetoda}).
A direct proof of this also follows by utilising the Lie algebraic techniques
described in Section 2 of ref. \cite{otu}.


\subsection{The $N=2$ Super Toda Solutions}

To begin with, we note that the $N=2$ WZNW equations of motion
(\ref{twotwoeqm})
are
solved by
\begin{eqnarray}\label{okfine}
g(x,t) &=& u(x^{++})W(A)v(x^{--}), \qquad  W(A)
=\exp(\int_{x^{--}}^{x^{++}}\!\!
A(s)\, ds),
\nonumber\\
\psi_+^a(x,t) &=& \psi_L^a(x^{++}),
\quad \bar\psi_+^{\bar a}(x,t) =
                     \bar\psi_L^{\bar a}(x^{++}),\quad \nonumber\\
\psi_-^a(x,t) &=& \psi_R^a(x^{--}),
\quad \bar\psi_-^{\bar a}(x,t) = \bar\psi_R^{\bar a}(x^{--}).\quad
\end{eqnarray}
Inserting these into the constraints (\ref{twotwofcon}-\ref{twotwosusypart}),
gives
\begin{eqnarray}\label{wellthatsgreat}
&& \psi_L|_- = \kappa\mu^aE^a_-, \qquad \psi_R|_+ = \kappa\nu^aE^a_+,
\nonumber\\
&& \bar\psi_L|_- = \kappa\mu^{\bar a}E^{\bar a}_-,
\qquad \psi_R|_+ = \kappa\nu^{\bar a}E^{\bar a}_+, \nonumber\\
&&\psi_L^2|_- = 0 = \bar\psi_L^2|_-,
\nonumber\\
&&\psi_R^2|_+ = 0 = \bar\psi_R^2|_+,
\nonumber\\
&&(\partial_{++}u\, u^{-1} + u A(x^{++})u^{-1} + \{\psi_L,\bar\psi_L\} )
|_-
           = 0, \nonumber\\
&&(v^{-1}\partial_{--}v  - v^{-1} A(x^{--})v - \{\psi_R,\bar\psi_R\} )|_-
           = 0.
\end{eqnarray}
Now we perform Gauss decompositions as in the $N=1$ supersymmetric
case. Writing $u=d_Lb_Lc_L$ and $v=d_Rb_Rc_R$, the bosonic
constraints may be shown
to reduce to
\begin{eqnarray}\label{fabulous}
(\partial_{++}c_L\, c_L^{-1} + c_L A(x^{++})c_L^{-1})|_- &=&
  -(b_L^{-1}d_L^{-1}\{\psi_L,\bar\psi_L\}d_Lb_L)|_-
           , \nonumber\\
(d_R^{-1}\partial_{--}d_R  - d_R^{-1} A(x^{--})d_R)|_- &=&
(b_Rc_R\{\psi_R,\bar\psi_R\}c_R^{-1}b_R^{-1})|_-.
\end{eqnarray}
We now fix the gauge symmetries of the parameterisations and constraints,
in a similar way to
the $N=1$ case. With regard to the fermionic constraints in equation
(\ref{wellthatsgreat}),
these are easily solved.
Then we fix $u=1$ (so that $c_L=b_L=d_L=1$) by
the gauge symmetry in the solution (\ref{okfine}). Next we set to zero the
fields
$\bar\psi_L^{\bar a},
\psi_L^a$
($\bar\psi_R^{\bar a},
\psi_R^a$), for $\alpha_{\bar a,a}$ the positive (negative of the) simple
roots,
 as these fields do not
appear in the constraints.
We similarily set the components of $\psi_L, \bar\psi_L$ along the directions
of
the negative non-simple roots to zero, and the components of $\psi_R,
\bar\psi_R$
along the directions of the positive non-simple roots to zero.
The second equation in equation (\ref{fabulous})
above has a triangular
symmetry whereby one may set $d_R=1$. Finally, on the right-hand side of the
second equation in equation (\ref{fabulous})
above, if one considers the coefficients of the
fermionic generators corresponding to the positive roots, one finds
a shift symmetry which can be used to gauge away the only piece of $c_R$ which
contributes to this expression. Thus one may set $c_R=1$ in this equation.

Thus we come to the following solution of the constraints
(\ref{wellthatsgreat})
\begin{eqnarray}\label{sssolution}
\psi_L &=& \kappa\mu^a E^a_- + \xi_L^iH^i,\nonumber\\
\bar\psi_L &=& \kappa\mu^{\bar a} E^{\bar a}_- + \xi_L^{\bar i}H^{\bar i},
\nonumber\\
\psi_R &=& \kappa\nu^a E^a_+ + \xi_R^iH^i,\nonumber\\
\bar\psi_R &=& \kappa\nu^{\bar a} E^{\bar a}_+ + \xi_R^{\bar i}H^{\bar i},
\nonumber\\
A|_{H} &=& a^iH^i + a^{\bar i} H^{\bar i}, \nonumber\\
A|_{a,\bar a<0} &=& -\{\psi_L,\bar\psi_L\}|_{a,\bar a<0}, \nonumber\\
A|_{a,\bar a>0} &=& (b_R\{\psi_R,\bar\psi_R\}b_R^{-1})|_{a,\bar a>0},
\end{eqnarray}
with the other components of $A$ equal to zero. \lq$|_{a,\bar a>0}$' means the
components along the directions of $E_+^{\bar a}, E_+^a$, and similarily
\lq$|_{a,\bar a<0}$' means the
components along the directions of $E_-^{\bar a}, E_-^a$.
The \lq\lq coordinates'' of this parameterisation of the constraint solution
are the fields $\xi_R^i(x^{--})$, $\xi_R^{\bar i}(x^{--})$, $\xi_L^i(x^{++})$,
$\xi_L^{\bar i}(x^{++})$, $a^i(s)$, $a^{\bar i}(s)$, $\phi_R^i(x^{--})$ and
  $\phi_R^{\bar i}(x^{--})$.
We may reconstruct the expression for the field $g$ using
$g=W(\hat A)\exp(\phi_R\cdot H)$, where the field $\hat A$ in
$W(\hat A)$ has non-zero
components given by the expressions for the components of $A$
in equation (\ref{sssolution}).

The explicit expressions for the Toda fields in terms of the solution
parameters are then
\begin{eqnarray}\label{todasman}
\phi^i(x,t) &=& \phi_R^i(x^{--}) - {\rm log}\lla W(\hat A)\lra, \nonumber\\
\bar\phi^{\bar i}(x,t) &=& \phi_R^{\bar i}(x^{--}) - {\rm log}\llab
             W(\hat A) \lrab, \nonumber\\
\chi^i_+(x,t) &=&
 {\lla (\kappa\mu^aE^a_- + \xi^j_LH^j)W(\hat A)\lra\over
                    \lla W(\hat A)\lra},\nonumber\\
\chi^{\bar i}_+(x,t) &=&
 {\llab (\kappa\mu^{\bar a}E^{\bar a}_- +
\xi^{\bar j}_LH^{\bar j})W(\hat A)\lrab\over
                    \llab W(\hat A)\lrab},\nonumber\\
\chi^i_-(x,t) &=&
 {\lla W(\hat A) (\kappa\exp(K_{a\bar b}\bar\phi_R^{\bar b})\nu^aE^a_+
        + \xi^i_RH^i)|\lra\over
                    \lla W(\hat A)\lra},\nonumber\\
\chi^{\bar i}_-(x,t) &=&
 {\llab W(\hat A) (\kappa\exp(K_{\bar a b}\phi_R^b)\nu^{\bar a}E^{\bar a}_+
       +\xi^{\bar i}_RH^{\bar i})\lrab\over
                    \llab W(\hat A)\lrab}.
\end{eqnarray}
These fields satisfy the $N=2$ Toda field equations
\begin{eqnarray}\label{todafieldeqns}
\partial_{++}\partial_{--}\phi^i &=& \alpha_i(K_{i\bar j}\chi_-^{\bar j})
                    (K_{i\bar k}\chi^{\bar k}_+)
                         \exp(K_{i\bar l}\bar\phi^{\bar l})
                    \nonumber\\
    & &\quad\quad       -\alpha_i\alpha_{\bar j}
                        K_{i\bar{j}}\exp(K_{\bar{j}l}\phi^l +
K_{i\bar k}\bar\phi^{\bar k}),
                        \nonumber\\
\partial_{--}\chi_+^i &=& \alpha_i(K_{i\bar j}\chi_-^{\bar j})
                                \exp(K_{i\bar k}\bar\phi^{\bar k}),
                               \nonumber\\
\partial_{++}\chi_-^i &=& -\alpha_i(K_{i\bar j}\chi_+^{\bar j})
                                \exp(K_{i\bar k}\bar\phi^{\bar k}),
\end{eqnarray}
and their conjugates (which arise from interchanging barred and
unbarred quantities). In the above, $\alpha_i=-\kappa^2\mu^i\nu^i$.
We may write equation (\ref{todafieldeqns}) in $N=2$ superspace as follows.
Define $D_+ = {\partial\over\partial\theta^+} + \bar\theta^+\partial_{++}$
and
$D_- = {\partial\over\partial\theta^-} + \bar\theta^-\partial_{--}$, and
their conjugates.
Introduce superfields
$\Phi^i$ and $\bar\Phi^{\bar i}$
satisfying the chirality constraints $D_{\pm}\bar\Phi^{\bar i}=0$
and $\bar D_{\pm}\Phi^i=0$.
Then make the component-field definitions
$\phi^i = \Phi^i|_0, \chi^i_\pm = D_\pm
\Phi^i|_0$
and the conjugates of these equations.
Then the equations of motion (\ref{todafieldeqns})
may be written in $N=2$ superspace as
\begin{equation}\label{superfieldeqns}
D_+D_-\Phi^i = \alpha_i\exp(K_{i\bar j}\bar\Phi^{\bar j}),
\end{equation}
(and its conjugate) where the
auxiliary fields are eliminated using their
non-propagating equations of motion.


\section{The Phase Spaces of Super Toda Theory}

Well-defined Hamiltonian formulations of the super WZNW models will
reduce under (first class) constraints
to well-defined Hamiltonian formulations of the super Toda theories.
Utilising the space of solutions as the phase space we may thus use
the super Toda solutions
derived in the previous section to parameterise the super Toda phase spaces.
It remains to give the Poisson brackets for the variables defining
the solutions, which we will now do.

The symplectic form for the $N=1$ super Toda theory is
\beq\label{symoneone}
\Omega = \half\int dx\left(\int d\theta^+(K_{ij}\delta\Phi^iD_+
\delta\Phi^j)\vert_{\theta^-=0}
 - \int d\theta^-(K_{ij}\delta\Phi^iD_-\delta\Phi^j)\vert_{\theta^+=0}\right).
\eeq
One can check that this is closed and time independent, using the equations
of motion (\ref{toda}). This is most directly done in components.
Now we insert the solution (\ref{bigres}) (the complexity of this
calculation is much reduced by using the time independence of the symplectic
form
to set $t=0$ in the Toda solutions and their derivatives).
This leads to
\beq\label{wonder}
\Omega = \int dx K_{ij} (\delta \phi_R^i\delta(\phi_R^j+2a^j) +
\half\delta\psi_+^i
               \delta\psi_+^j   -\half\delta\psi_-^i
               \delta\psi_-^j).
\eeq
Thus the Poisson brackets become those of the non-supersymmetric Toda
model, with parameters $(\phi_R^i,a^i)$ ({\it c.f.} ref. \cite{pstoda}),
together with free fermion Poisson
brackets for the fields $\psi^i_\pm$.

The $N=2$ case is similar to the $N=1$ case just presented.
The symplectic form of the $N=2$ Toda theory is
\begin{equation}\label{symplecticform}
\Omega = \int \! dx\, K_{i\bar j}\left(\delta\phi^i\delta\dot{\bar\phi^{\bar
j}}
                + \delta\bar\phi^{\bar j}\delta\dot\phi^i + \delta\chi^i_+
              \delta\chi_+^{\bar j} - \delta\chi_-^i\delta\chi^{\bar j}_-
                    \right).
\end{equation}
One can check that this is closed and time independent, using the Toda
equations of motion. It is also straightforward to check that it is
$N=2$ supersymmetric - for example, one of the right supersymmetries is
given by the transformations $\Delta\phi^i=\epsilon^+\chi_+^i,
\Delta\chi_+^{\bar i} = \epsilon^+\partial_{++}\bar\phi^{\bar i}$,
with the other variations zero, and one can show directly that
$\Delta\Omega=0$.
Inserting the Toda solutions (\ref{todasman})
into the symplectic form (\ref{symplecticform})
gives in a straightforward way
\begin{equation}\label{symplecticformtwo}
\Omega = \int \! dx\, K_{i\bar j}\left(\delta\phi_R^i\delta(\bar\phi_R^{\bar j}
                       +2a^{\bar j})
                + \delta\bar\phi_R^{\bar j}\delta(\phi_R^i +2a^i)
                     + \delta\xi^i_+
              \delta\xi_+^{\bar j} - \delta\xi_-^i\delta\xi^{\bar j}_-
                    \right).
\end{equation}
Thus we see again the Poisson brackets of the bosonic model together with
free field brackets for the fermions. In the $N=1$ and $N=2$ cases we see from
equations (\ref{wonder}) and (\ref{symplecticformtwo}) that the solution
parameter variables are simply related to the usual canonical Hamiltonian
variables (the initial data).


\section{Concluding Remarks}

In this paper we have presented new formulations of the solutions of
the $N=1$ and $N=2$ supersymmetric Toda theories. These were obtained
using Hamiltonian reduction on formulations of the solutions of the
corresponding WZNW models which we presented in a previous paper. The
advantage of our formulations of the super Toda solutions is that
they allow a direct correspondence between the solutions and the
initial data. Using this, one can then formulate the
phase space description using the space of solutions, which can then
be seen to be equivalent to the Hamiltonian formulation.

Although we have not presented details here,
these methods can also be applied directly to other Toda systems - the
various models obtained by non-standard reductions, and the models obtained by
reductions of the affine WZNW theories, for example.


\section{Acknowledgements}
GA was supported by an Australian Postgraduate Research Award,
and BS by a QEII Fellowship from the Australian Research Council, and currently
by the Engineering and Physical Sciences Research Council of the UK.

\end{document}